# Atomic insight into Li$^+$ ion transport in amorphous electrolytes Li$_x$AlO$_y$Cl$_{3+x-2y}$ (0.5 ≤ x ≤ 1.5, 0.25 ≤ y ≤ 0.75)

Qifan Yang[1, 2], Jing Xu[1, 3], Xiao Fu[1, 2], Jingchen Lian[1, 2], Liqi Wang[1, 3], Xuhe Gong[1, 4], Ruijuan Xiao[1, 2, 3], * and Hong Li[1, 2, 3], *

[1] Beijing National Laboratory for Condensed Matter Physics, Institute of Physics, Chinese Academy of Sciences, Beijing 100190, China

[2] Center of Materials Science and Optoelectronics Engineering, University of Chinese Academy of Sciences, Beijing 100049, China

[3] School of Physical Sciences, University of Chinese Academy of Sciences, Beijing 100049, China

[4] School of Materials Science and Engineering, Key Laboratory of Aerospace Materials and Performance (Ministry of Education), Beihang University, Beijing 100191, China

E-mail: rjxiao@iphy.ac.cn, hli@iphy.ac.cn

**Abstract:** The recent study of viscoelastic amorphous oxychloride electrolytes has opened up a new field of research for solid-state electrolytes. In this work, we chose Li-Al-O-Cl system containing disordered structures with varying O/Cl ratio and Li$^+$ content to study their structural characteristics and ion transport mechanism using ab-initio molecular dynamics (AIMD) simulation and machine learning interatomic potential based molecular dynamics (MLIP-based MD) simulation. It is found that O-doping results in the presence of a skeleton of Al-chains formed by AlOCl tetrahedra and the increase of glass forming ability, causing Cl atoms' rotation around centered-Al within the tetrahedron thus facilitating the motion of Li$^+$ ions. However, further increase of O/Cl ratio decreases the number of rotating Cl atoms, weakening the transport of Li$^+$. So increasing glass forming ability without reducing Cl content or by methods through controlling synthesis conditions, are useful to promote Li$^+$ ion conducting of oxychloride electrolytes.

# 1 Introduction

At present, the fast development of lithium-ion batteries (LIBs) with excellent properties[1] has promoted the progress of industries[2, 3]. Compared to the commonly used LIBs, solid-state batteries (SSBs) with solid-state electrolytes (SSEs) have better safety and greater potential[4] - [7]. SSEs can be divided into polymer SSEs, inorganic SSEs and their composites[8, 9]. Among them, inorganic SSEs consist of two structural types: crystalline and amorphous[10, 11]. Up to now, crystalline SSEs have been well explored. Li Argyodites[12] - [14], $Li_{10}GeP_2S_{12}$ (LGPS)[15, 16], NASICON structure of $LiM_2(PO4)_3$ (M = Ge, Ti, Sn, Hf, Zr) compositions[17], garnet structure of $Li_xLa_3M_2O_{12}$ (5 ≤ $x$ ≤ 7, M = Nb, Ta, Sb, Zr, Sn) compositions[18, 19] and so on exhibit ionic conductivity similar to that of liquid electrolytes. Compared to crystalline SSEs, there is much less exploration of amorphous SSEs due to the complexity of their structures[20]. However, amorphous SSEs are also likely to exhibit high ionic conductivity[21], for example, the $Li_2S-P_2S_5$ glass system has been extensively studied as SSEs[22]. Previously studied amorphous SSEs mostly belong to sulfides or oxides, but with the emergence of crystalline halide electrolytes[23], some amorphous oxychloride electrolytes with good conductivity and compressibility began to appear. Hu et al.[24] synthesized partially amorphous oxychloride $Li_{1.75}ZrCl_{4.75}O_{0.5}$ which shows an ionic conductivity of 2.42 mS cm$^{-1}$ at room temperature. Zhang et al.[25] presented amorphous SSEs, $xLi_2O-MCl_y$ (M = Ta or Hf, 0.8 ≤ x ≤ 2, y = 5 or 4) possessing ionic conductivities up to 6.6 mS cm$^{-1}$ at 25 °C. In addition, Dai et al.[26] reported glasses $MAlO_xCl_{4-2x}$ (MAOC, M = Li, Na, 0.5 < x < 1) by doping O atoms in $LiAlCl_4$, and the ionic conductivity at room temperature of $Li_{1.0}AlO_{0.75}Cl_{2.5}$ reached ~1 mS cm$^{-1}$. The first-principles calculations on $MAlO_xCl_{4-2x}$ indicated that Al atoms are connected into chains through O or Cl, forming a Al chain skeleton in these structures and the motion of Li and Cl have correlation[26]. Although the preliminary understanding of amorphous SSEs has been formed in our previous studies[26, 27], it is still unclear about the formation mode of the disordered atomic arrangement, the mechanism and characteristics of Li$^+$ ion transport, as well as the effect of O/Cl ratio on atomic structure and ionic conductivity. It is necessary to further analyze this system to figure out above questions and try to gain the design criterion for the amorphous SSEs.

To deepen the understanding of amorphous solid electrolytes, we generate and simulate the disorder Li-Al-O-Cl system in various compositions through ab-initio molecular dynamics (AIMD), which is a relatively accurate method commonly used for crystalline systems in the simulation of structural evolution and ionic transport[28, 29]. However, compared to crystals, computational simulation by AIMD has some unique challenges for amorphous materials. Firstly, the atomic arrangement of amorphous materials exhibits long-range disorder without translational periodicity symmetry, which makes it difficult to establish atomic models for simulation. Typically, the construction of the amorphous model in AIMD simulation is performed in a fictitious cubic box through melting and rapid quenching process to obtain the disordered arrangement of atoms[30]. Due to the large computational cost of this method, the systems that are adopted in simulations are generally smaller than 200 atoms, corresponding to a lattice length of 10~20 Å. The size effect of the model and the accessible time scale for evolution become the main limitations on simulation effectiveness. Secondly, the room-temperature ionic conductivity in amorphous configurations cannot be extrapolated from high-temperature results as crystals, as amorphous configurations constantly relax, leading to different structures at different temperatures. Direct simulations at room temperature are only possible for systems with high ion transport capabilities, as weak transport cannot generate enough migration events in a limited simulation time to obtain accurate statistical results. Thus although AIMD can accurately illustrate each conducting event and ion transport mechanism[31] - [33], as well as provide us with initial configurations of disordered structures and simulate ion migration within a certain range, research on larger-spatial and longer-time scales and obtaining macroscopic overall ionic conductivity requires faster computational methods, such as recently developed machine learning interatomic potential based molecular dynamics (MLIP-based MD).

In this work, five different compositions for amorphous LAOCs ($Li_{0.5}AlO_{0.75}Cl_2$, $Li_{1.0}AlO_{0.75}Cl_{2.5}$, $Li_{1.5}AlO_{0.75}Cl_3$, $Li_{1.0}AlO_{0.25}Cl_{3.5}$ and $Li_{1.0}AlCl_4$) were simulated through AIMD method and MLIP method to analyze the atomic structural and ion conducting characteristics in larger and longer scales, identify the physical origin of the excellent $Li^+$ transport performance of amorphous oxychloride, and present the possible strategies to explore amorphous SSEs.

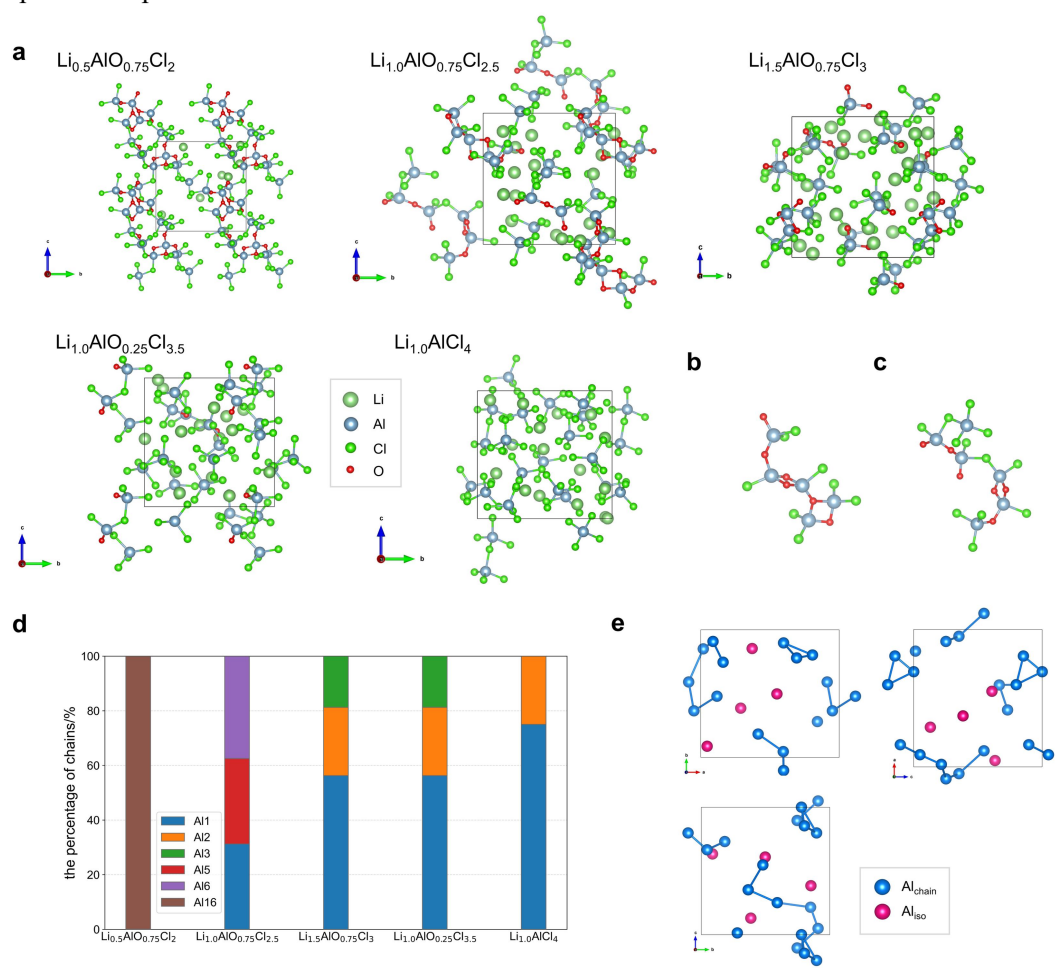

**Figure 1.** Structural analysis of LAOCs calculated by AIMD. (a) The structures of amorphous LAOCs in five compositions. (b) Al5 and (c) Al6 chains found in $Li_{1.0}AlO_{0.75}Cl_{2.5}$ structure. (d) The percentage of various Al chains in each of the five structures (For example, Al16 refers to a Al chain of 16 Al atoms). (e) The distribution of $Al_{chain}$ and $Al_{iso}$ in $Li_{1.0}AlO_{0.75}Cl_{2.5}$ structure viewed in different directions.

# 2 Results and discussion

**2.1 Structural characteristics in LAOCs**

The initial atomic structures for amorphous LAOCs are built through AIMD simulations for melting and quenching process, for which the detailed method and parameters are described in Supporting Information. The most significant feature of amorphous structures is the Al-chains formed by corner- or edge-sharing tetrahedra as shown in Figure 1a. For comparison, the crystalline LiAlCl$_4$ is illustrated in Figure S1. In addition, the change of Al-chains length during AIMD simulation of Li$_{1.0}$AlO$_{0.75}$Cl$_{2.5}$ is shown in Figure S2.

Li$_{1.0}$AlO$_{0.75}$Cl$_{2.5}$ is used to illustrate the structural characteristics of amorphous LAOCs. As shown in Figure 1a, the structure contains long chains consisting of AlO$_m$Cl$_{4-m}$ (m=1,2,3,4) tetrahedra (AOC), and the Al atom on the chains is labeled as Al$_{chain}$. The isolated AOC tetrahedra also exist, and the Al atom in the center of isolated AOC tetrahedra is marked as Al$_{iso}$. In Li$_{1.0}$AlO$_{0.75}$Cl$_{2.5}$, two types of Al atom chains are found according to the number of Al atoms in the chain, which are Al5 and Al6 illustrated in Figure 1b and 1c. As we can see, Al atoms are connected by Cl or O atoms in the structure to form Al chains, and when the two adjacent Al$_{chain}$ are connected by Cl atoms, the distance between them reaches the maximum not exceeding 4.2Å. Therefore, in the analysis for other compositions, the Al atoms with distances less than 4.2Å are considered to belong to the same chain. The statistical results on the percentage of Al chains with different chain lengths for each LAOCs were presented in Figure 1d. It shows that Al atoms in Li$_{0.5}$AlO$_{0.75}$Cl$_2$ are all Al$_{chain}$, forming Al16 chains. This chain contains 16 Al atoms, and the AOC tetrahedra are mostly connected in an Al-O-Al form, indicating that the formation of these long chains may be related to the high oxygen content in structure. As the O/Cl ratio decreases, the number of Al$_{chain}$ gradually decreases and the number of Al$_{iso}$ increases, and when the doping concentration of O decreases to 0, the amorphous Li$_{1.0}$AlCl$_4$ contains the most Al$_{iso}$. It is noted that Li$_{1.5}$AlO$_{0.75}$Cl$_3$ and Li$_{1.0}$AlO$_{0.25}$Cl$_{3.5}$ have the same proportion of Al$_{chain}$ and Al$_{iso}$, possibly originating from the limited simulation cells used in AIMD and the larger simulation box may exhibit more abundant structural details, thus better distinguishing these two structures. The structural characteristics also reflect in the radial distribution function (RDF) of Al-Al pair in Figure S3, S4. The first coordination in the Al-Al RDF curve of amorphous LAOCs originates from two adjacent Al$_{chain}$, since r (distance between Al-Al) is between 2-4 Å which is exactly the range for distances between adjacent Al$_{chain}$. As the O/Cl ratio decreases, the first coordination number gradually decreases, indicating the number of Al$_{iso}$ increases and the number of Al$_{chain}$ decreases. The findings above indicate that the length and proportion of Al chains in the LAOC structure can be controlled by adjusting the O/Cl ratio. Finally, as Figure 1e shows, Al$_{chain}$ and Al$_{iso}$ in Li$_{1.0}$AlO$_{0.75}$Cl$_{2.5}$ are distributed in a random way with no segregation phenomenon, which means the structure of LAOC is more similar to polymers formed by chains of various lengths.

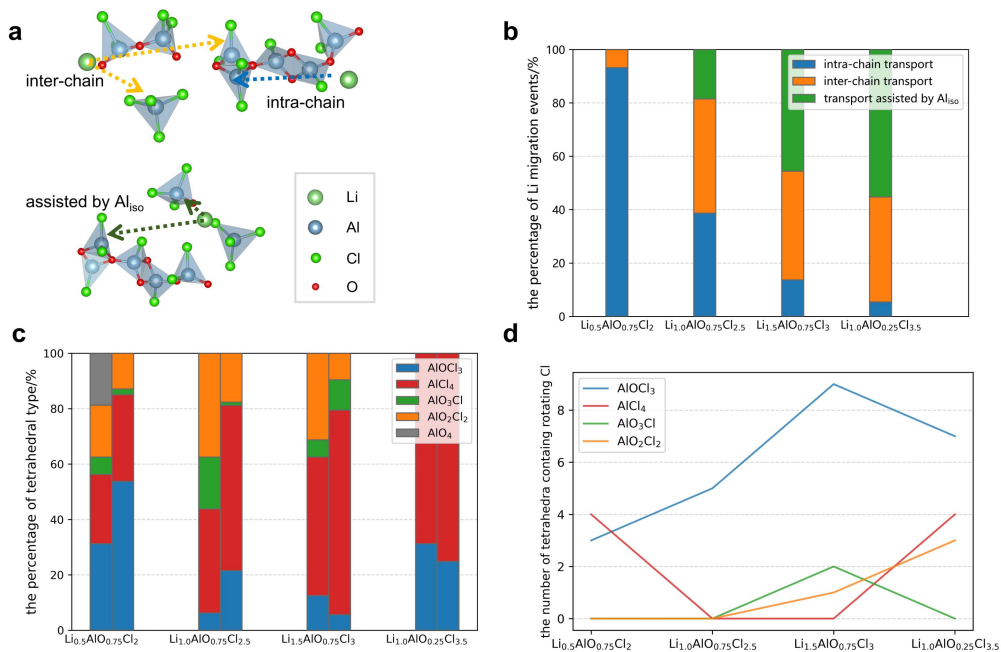

**Figure 2.** Analysis of ion transport events in LAOCs calculated by AIMD. (a) Schematic diagram of intra-chain transport event, inter-chain transport event and transport event assisted by $Al_{iso}$. (b) The percentage of three different transport events in LAOCs in 300K's AIMD simulations within the range of 11-50ps. (c) For each composition, left column: the percentage of the five tetrahedra type in the original structure of LAOCs; right column: the percentage of tetrahedra types located nearest to the migrating $Li^+$ ion in all transport events in LAOCs. (d) The number of tetrahedra types with rotating Cl atom in LAOCs.

**2.2 Ionic transport mechanism of LAOCs**

Due to that the local environment of Al in the LAOC structure are classified into $Al_{chain}$ and $Al_{iso}$, the transport events of $Li^+$ ions can be divided into intra-chain transport, inter-chain transport, and transport assisted by $Al_{iso}$. In these structures, $Li^+$ ions are connected to the Al-chains through either $Cl^-$ or $O^{2-}$. Intra-chain transport refers to the migration of $Li^+$ from one AOC tetrahedron to another one in the same chain and inter-chain transport is the migration of $Li^+$ from a site bonding to one chain to the adjacent site connecting to another chain or another isolated AOC tetrahedron. Transport assisted by isolated Al refers to the migration of Li from one isolated AOC tetrahedron to another local environment. (Figure 2a).

We conduct a statistical analysis by calculating the proportions of $Li^+$ transport events and their nearest local environment of LAOCs in 11-50 ps's AIMD simulation. The percentage of three different transport events of each LAOC composition at 300K was displayed in Figure 2b. The results revealed that as the O/Cl ratio decreases, the intra-chain transport gradually decreases, while the inter-chain and $Al_{iso}$-assisted transport increase. This is related to the change in atomic structure, in which the length and number of Al chains are gradually decreasing. The chemical environments for all the $Li^+$ migration events are analyzed and the conditions benefit to $Li^+$ transport are extracted. Figure 2c indicates the percentage of various tetrahedra in the initial structure (left column) and the percentage of tetrahedral types related to $Li^+$ transport events (right column). It suggests that the majority of the tetrahedra near the migrating $Li^+$ are $AlCl_4$ and $AlOCl_3$ tetrahedra (over 75%), indicating that the migration of $Li^+$ ions is more likely to occur in localized environments with high chlorine content. Furthermore, according to the analysis of the structural features mentioned earlier, O atoms mainly serve as corner- or edge-sharing atoms to connect tetrahedra, hence a higher chlorine content implies a shorter Al chain length. Therefore, under the same conditions, the higher the sum of the tetrahedral ratios of $AlCl_4$ and $AlOCl_3$ in structures, the shorter the Al chains, the higher the ionic conductivity is expected to be found. While $Li^+$ migrates in the disordered structures, for Cl atoms, unlike remaining stably unmoved in crystal $LiAlCl_4$ (Figure S5), some Cl atoms also move in LAOCs, as shown in Figure 3c and Figure S6, the MSD of Cl is quite large. Different from Li, the Cl atom mostly rotates around the central Al of the AOC tetrahedron, without undergoing long distance migration. The changes in the Al-Cl bond lengths throughout the entire simulation process are shown in Figure S7, indicating that the Al-Cl bonds are not broken, demonstrating that Cl moves by rotation in its AOC tetrahedron. The only exception is the migration of one Cl atom connected with $AlOCl_3$ tetrahedron in $Li_{1.0}AlO_{0.25}Cl_{3.5}$. Figure 2d counted the tetrahedra with rotating Cl in different compositions proving that Cl movement on $AlCl_4$ and $AlOCl_3$ is the most intense, indicating that the rotation of Cl on $AlCl_4$ as well as $AlOCl_3$ and $Li^+$ transport can promote each other.

The above analysis indicates that the rotation of Cl within AOC tetrahedra of LAOC structures promotes the transport of $Li^+$ ions. Compared with $AlO_2Cl_2$, $AlO_3Cl$, and $AlO_4$, the proportion of stable O atoms on the tetrahedra of $AlCl_4$ and $AlOCl_3$ is smaller, and the inhibitory effect on the rotation of Cl connected to the same Al atom is also smaller. At the same time, the proportion of Cl atoms is higher, so once a Cl rotates, it is likely to drive the rotation of the surrounding Cl atoms. Therefore, the rotation of Cl on $AlCl_4$ and $AlOCl_3$ is much more intense than in others. When the ratio of $AlCl_4$ and $AlOCl_3$ is high, it can greatly promote the movement of $Li^+$. Therefore, in amorphous LAOCs, under the same Li and Al concentrations, the lower the O/Cl ratio (the more Cl atoms), the more intense $Li^+$ transport, and the higher expected $Li^+$ conductivity.

To prove the aforementioned conclusion, amorphous $Li_{1.0}AlCl_4$, $Li_{1.0}AlO_{0.25}Cl_{3.5}$ and $Li_{1.0}AlO_{0.75}Cl_{2.5}$, with the same concentration of Li and Al, were selected. According to above analysis, the ion conductivity performance among these structures should be: $Li_{1.0}AlCl_4$ > $Li_{1.0}AlO_{0.25}Cl_{3.5}$ > $Li_{1.0}AlO_{0.75}Cl_{2.5}$. To confirm the inference, the AIMD simulations were conducted for these systems at 300K, from which the statistically obtained MSDs are shown in Figure 3b. The MSD of the three structures is ranked as $Li_{1.0}AlCl_4$ > $Li_{1.0}AlO_{0.25}Cl_{3.5}$ > $Li_{1.0}AlO_{0.75}Cl_{2.5}$, which further confirms the correctness of above conclusions.

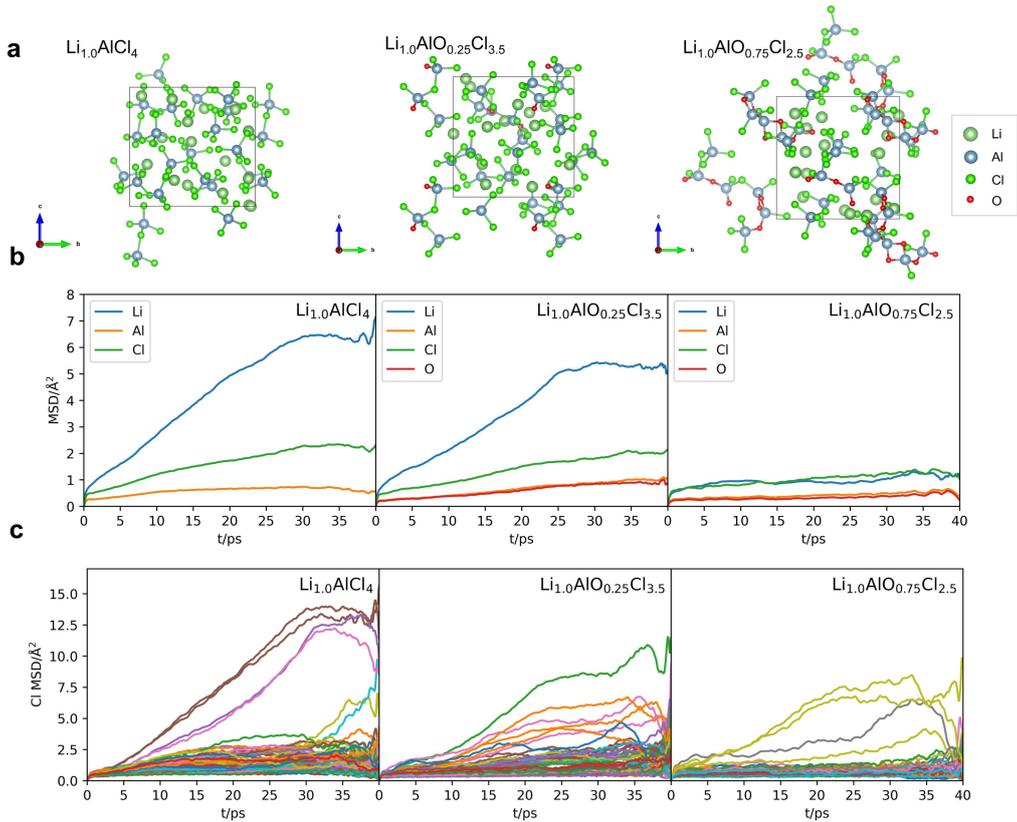

**Figure 3.** Structures and MSDs of amorphous $Li_{1.0}AlCl_4$, $Li_{1.0}AlO_{0.25}Cl_{3.5}$ and $Li_{1.0}AlO_{0.75}Cl_{2.5}$ calculated by AIMD at 300 K. (a) The simulation models of $Li_{1.0}AlCl_4$, $Li_{1.0}AlO_{0.25}Cl_{3.5}$ and $Li_{1.0}AlO_{0.75}Cl_{2.5}$. (b) MSDs for $Li_{1.0}AlCl_4$, $Li_{1.0}AlO_{0.25}Cl_{3.5}$ and $Li_{1.0}AlO_{0.75}Cl_{2.5}$ at 300K. (c) MSD of each Cl atom for $Li_{1.0}AlCl_4$, $Li_{1.0}AlO_{0.25}Cl_{3.5}$ and $Li_{1.0}AlO_{0.75}Cl_{2.5}$ at 300K.

Based on above analysis, it is inferred that the doping effect of O in LAOC on $Li^+$ ion transport is a game of two factors. The addition of element O enhances the system's ability to form amorphous structures, promoting the rotation of Cl and the transport of Li. However, the decrease in Cl content reduces the number of $AlCl_4$ and $AlOCl_3$ tetrahedra, which has a negative effect on the transport performance of $Li^+$ ions. This also indicates that besides O doping, if we can find other ways to improve the amorphization degree of $LiAlCl_4$, it may be possible to achieve higher $Li^+$ ion conductivity at higher Cl content.

## 2.3 Machine learning interatomic potential based molecular dynamics (MLIP-based MD) simulation under large supercells

In AIMD simulations, the MSDs of lithium ions at 300K in Figure 3b is relatively small (<10Å$^2$), indicating very few migration events which makes it difficult to evaluate the ion conductivity. Therefore, a larger cell and a longer period of simulation time are needed. Using the AIMD data as training set, we obtained the MLIP for Li-Al-O-Cl quaternary system through DeePMD method[32], which is one of the efficient softwares for training MLIP to accelerate the dynamic simulation process. The detailed training process is described in Supporting Information. By adopting the MLIP-based MD, a large number of migration events occur during a longer simulation time and the ionic transport properties at 300K are estimated. Note S1 provides the the accuracy validation of the MLIP model and the study of size effects used to determine the cell size for subsequent MLIP-based MD simulations.

To build the amorphous model in larger boxes such as 2×2×2 and 3×3×3 supercell expanded from the initial box for $Li_{1.0}AlO_{0.75}Cl_{2.5}$, two different modelling strategies are tested. Although both models have the same initial structure, which is the supercell expanded from the original disordered cell obtained from AIMD simulations, the first method involves a process of re-melting and quenching then equilibrating by MLIP-based MD, while the second method directly conducts simulations at 300K on the expanded structure using MLIP-based MD. Note S2 introduces a detailed description of these two methods. Figure 4 illustrates the simulations' results for these two models in 3×3×3 expanded box. Figure 4a and 4b show that two methods yield very similar amorphous structures in $Li_{1.0}AlO_{0.75}Cl_{2.5}$, with a consistent distribution of the number of Al-chains and the proportion of AOC groups. However, model 1 has less $Al_{iso}$ and fewer numbers of the sum of $AlCl_4$ and $AlOCl_3$ groups, resulting in a slightly lower ionic conductivity statistically evaluated from the MSD than that of model two. The ionic conductivity of 12 mS/cm and 17 mS/cm was obtained for model 1 and model 2, respectively, indicating that due to the non-uniqueness of the disordered structure, the ionic conductivity of the system varies with the degree of disorder. Since the disorder of the structure in the simulation is much higher than that of the experimentally synthesized structure, the calculated conductivity is also higher than that obtained experimentally.

Therefore, we further investigated the relationship between $Li^+$ ion conductivity and degree of disorder. Since only $LiAlCl_4$ has a definite crystal structure in the studied system, we used $LiAlCl_4$ structures at different melting stages to manufacture different degrees of disorder in the non-crystalline $LiAlCl_4$ configurations. The AIMD simulations at 300K were carried out for these models with various amorphous degree. The results in Figure S8 indicate that in $LiAlCl_4$ more disordered structures show larger MSD values, which means higher $Li^+$ conductivity and more rotation events of Cl in AOC groups.

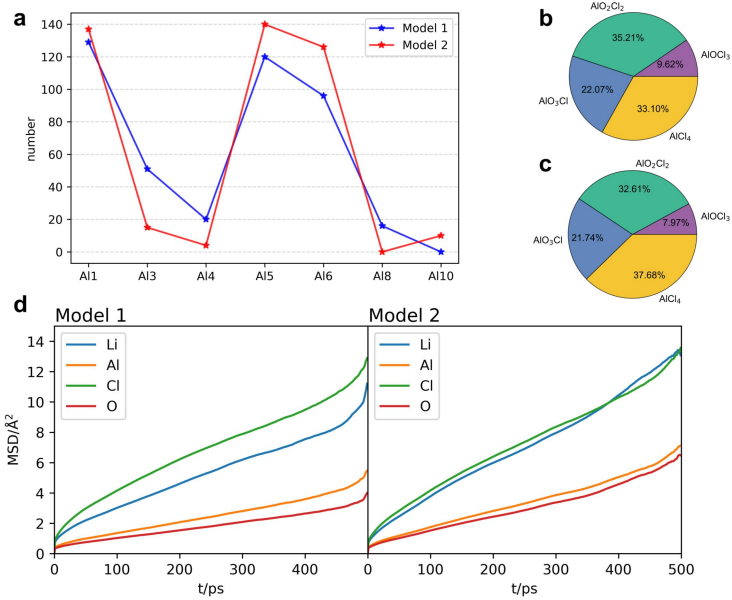

**Figure 4.** Comparison of model 1 and model 2 for $Li_{1.0}AlO_{0.75}Cl_{2.5}$ calculated by MLIP-based MD simulations. (a) The number of Al chains of different lengths in two models of $Li_{1.0}AlO_{0.75}Cl_{2.5}$. (b) The percentage of the four tetrahedra type of $Li_{1.0}AlO_{0.75}Cl_{2.5}$ in model 1. (c) The percentage of the four tetrahedra type of $Li_{1.0}AlO_{0.75}Cl_{2.5}$ in model 2. (d) MSDs at 300K for two models of $Li_{1.0}AlO_{0.75}Cl_{2.5}$.

## 3 Conclusions

By constructing the amorphous structure of $Li_xAlO_yCl_{3+x-2y}$ system and simulating the $Li^+$ ion conductivity properties, we have gained insights into the atomic structural features, ion transport mechanisms, as well as the modulation of ion conductivity by composition and structure in this type of disordered solid electrolyte. Based on this type of system, we have drawn the following conclusions. Firstly, in atomic structural analysis, through the O doping, the $AlCl_4$ groups originally existing isolatedly in $LiAlCl_4$ crystals transform into AOC groups, which are connected by O and Cl to form a skeleton of Al-chains with varying lengths. By adjusting the doping amount of O, the chain length distribution of Al-chains in the structure can be regulated. The main function of O doping is to enhance the system's glass forming ability, making it easier to synthesize amorphous structures in experiments. Secondly, in the analysis of ion transport performance, $Li^+$ ions migrate in the structure via both inter-chain and intra-chain modes, and the rotational freedom of Cl in the AOC groups enhances the migration ability of $Li^+$ ions. Due to the promoting effect of Cl rotation on $Li^+$ transport, the higher Cl/O ratio in the completely amorphous LAOC structure, the stronger the promoting effect on $Li^+$ motion, the better the ion transport performance of the structure. On the other hand, the doping of O helps to realize the amorphization of the LAOC structure, which is a prerequisite for Cl atoms to rotate and promote Li motion. Therefore, the doping of O in LAOC leads to the combined effect of these two factors, maximizing $Li^+$ ion conductivity at optimal doping concentrations. Besides, attempting other ways to increase the amorphization of chlorides without reducing Cl content is also interesting endeavors in the future. Thirdly, unlike in crystalline systems, the non-uniqueness of the amorphous structure prevents the interpolation of ion conductivity at room temperature or lower from high-temperature MD data, thus the practicality of MLIP-based MD simulations of amorphous solid state electrolytes was demonstrated and the ion conductivity was evaluated. Our simulations also indicate that even for structures with the same composition, the varying degrees of amorphization will result in different Li$^+$ ion conductivities. This provides an opportunity to adjust the amorphization degree of the system through control of synthesis conditions, thereby tuning the ionic conductivity performance.

**Author Contributions**

R. J. X. managed and designed the project, checked the manuscript and provided suggestions. Q. F. Y. performed all theoretical calculations and provided related explanations, drafted the manuscript, which underwent thorough revision with active participation and contributions from all authors. J. X., X. F, J. C. L., L. Q. W., X. H. G. provided suggestions for the theoretical calculations and manuscript.

**Conflicts of interest**

The authors declare no competing financial interest.

**ACKNOWLEDGMENTS**

This work was supported by funding from the National Natural Science Foundation of China (grants no. 52172258), and the Strategic Priority Research Program of Chinese Academy of Sciences (grant no. XDB0500200). The numerical calculations in this study were carried out on both the ORISE Supercomputer, and the National Supercomputer Center in Tianjin.

# Supporting Information for

# Atomic insight into Li$^+$ ion transport in amorphous electrolytes Li$_x$AlO$_y$Cl$_{3+x-2y}$ (0.5 ≤ x ≤ 1.5, 0.25 ≤ y ≤ 0.75)

Qifan Yang[1, 2], Jing Xu[1, 3], Xiao Fu[1, 2], Jingchen Lian[1, 2], Liqi Wang[1, 3], Xuhe Gong[1, 4], Ruijuan Xiao[1, 2, 3], * and Hong Li[1, 2, 3], *

[1] Beijing National Laboratory for Condensed Matter Physics, Institute of Physics, Chinese Academy of Sciences, Beijing 100190, China

[2] Center of Materials Science and Optoelectronics Engineering, University of Chinese Academy of Sciences, Beijing 100049, China

[3] School of Physical Sciences, University of Chinese Academy of Sciences, Beijing 100049, China

[4] School of Materials Science and Engineering, Key Laboratory of Aerospace Materials and Performance (Ministry of Education), Beihang University, Beijing 100191, China

E-mail: rjxiao@iphy.ac.cn, hli@iphy.ac.cn

**Methods**

**Structure acquisition:** Crystal structure of LiAlCl$_4$ was obtained from Materials Project database. Four different amorphous LAOCs (Li$_{0.5}$AlO$_{0.75}$Cl$_2$, Li$_{1.0}$AlO$_{0.75}$Cl$_{2.5}$, Li$_{1.5}$AlO$_{0.75}$Cl$_3$, Li$_{1.0}$AlO$_{0.25}$Cl$_{3.5}$) were obtained by doping with varying concentrations of O in the LiAlCl$_4$ crystal structure as well as removing corresponding amounts of Cl according to charge neutrality, and performing melting and quenching process by AIMD simulations. For comparison, the amorphous structure of Li$_{1.0}$AlCl$_4$ was also generated using the same method. Figure S3 and S4 prove the successful acquisition of amorphous LAOC structures. The structures were visualized by the VESTA software package[1].

**Ab initio molecular dynamics simulation:** We performed AIMD simulations to construct amorphous structural models and study the structure and ion transport characteristics of LAOCs (Table S1 shows the atom number of each cell) with lattice parameters larger than or near 10 Å, with nonspin-polarized DFT calculations using a Γ-centered k-point. The AIMD calculations was conducted with a Nose thermostat[2] for 30,000 steps at 600K (melting); 3,000 steps from 600K to 300K (quenching) and 50,000 steps at 300K (equilibrium) using a time step of 1 fs. The last configuration of the quenching process is adopted to perform the structural relaxation by density functional theory (DFT) computation to obtain the cell parameters through Vienna Ab initio Simulation Package (VASP)[3] and Perdew–Burke–Ernzerhof (PBE)[4], gradient approximation (GGA) by the projector-augmented-wave (PAW) approach. The cutoffs of the wavefunction and the density are 520 eV and 780 eV, respectively. Both cells and ions were relaxed to reach the energy and force convergence criterion of $10^{-5}$ eV and 0.01 eV/Å. For equilibrium process, the first 10 ps is used for equilibrating the structure, and the subsequent 40 ps is used for statistically obtaining the ion migration properties. We carried out all analysis by pymatgen and the pymatgen-diffusion Python packages.[5], [6]

**MLIP-based molecular dynamics simulation:** The MLIP model was trained by DeepMD[7] with the training set of AIMD simulation data of the original crystal structure of Li$_{1.0}$AlO$_{0.75}$Cl$_{2.5}$ each 30ps at 300K, 400K, 500K and 600K, using the descriptor "se_e2_a" and parameters derived from the examples provided by the DeePMD framework. The MD simulations of Li$_{1.0}$AlO$_{0.75}$Cl$_{2.5}$ by LAMMPS[8] (Large-scale Atomic/Molecular Massively Parallel Simulator) in original cell, 2×2×2 cell and 3×3×3 cell (Table S2 shows the atom number of each cell) were conducted at constant volume and temperature (NVT) ensemble. The MD simulation was carried out for 500,000 steps at 300K for original cell and 2×2×2 cell, with a time step of 0.1 fs. For 3×3×3 cell, Model 1 refers to the MD run for 3,000,000 steps (300 ps) at 600K (melting); 30,000 steps (3 ps) from 600K to 300K (quenching) and 5,000,000 steps (500 ps) at 300K (equilibrium), with a time step of 0.1 fs; Model 2 refers to the MD run for 5,000,000 steps (500 ps) at 300K, with a time step of 0.1 fs. For MD of Model 1 and 2 at 300K, the 0 ps to 500 ps is used for further analysis. The results of them were used to verify the accuracy of the model, study size effects and investigate structural and ion conducting characteristics in large boxes. We carried out all analysis by pymatgen and the pymatgen-diffusion Python packages. The open visualization tool (OVITO)[9] was used for the visualizing and analyzing of LAMMPS results.

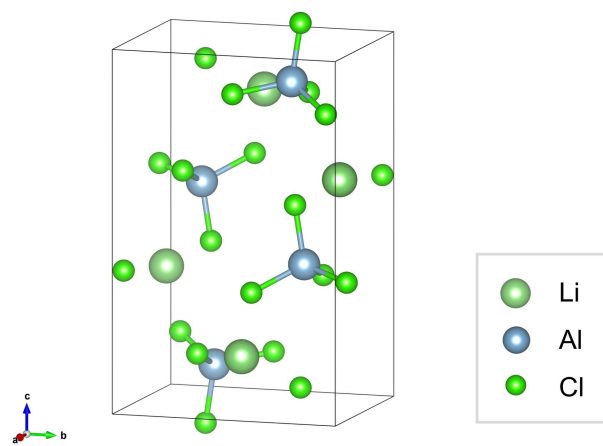

**Figure S1.** The structure of crystalline LiAlCl$_4$.

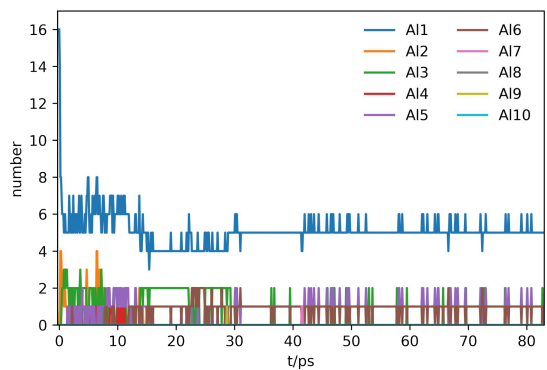

**Figure S2.** The variation of the number of Al-chains with different lengths over the whole AIMD simulation for Li$_{1.0}$AlO$_{0.75}$Cl$_{2.5}$ (30 ps at 600K (melting); 3 ps from 600K to 300K (quenching) and 50 ps at 300K (equilibrium).

It is found that the distribution of chain length changed basically only in the melting process.

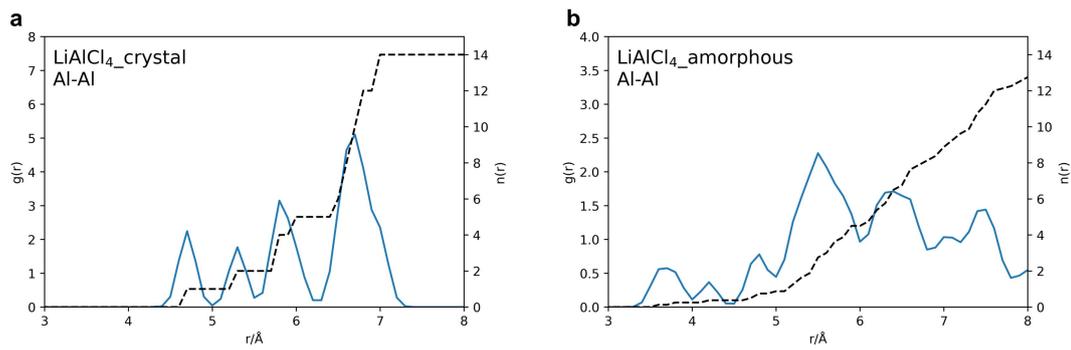

**Figure S3.** The RDF curves for Al-Al pairs of crystal and amorphous LiAlCl$_4$ at 300K.

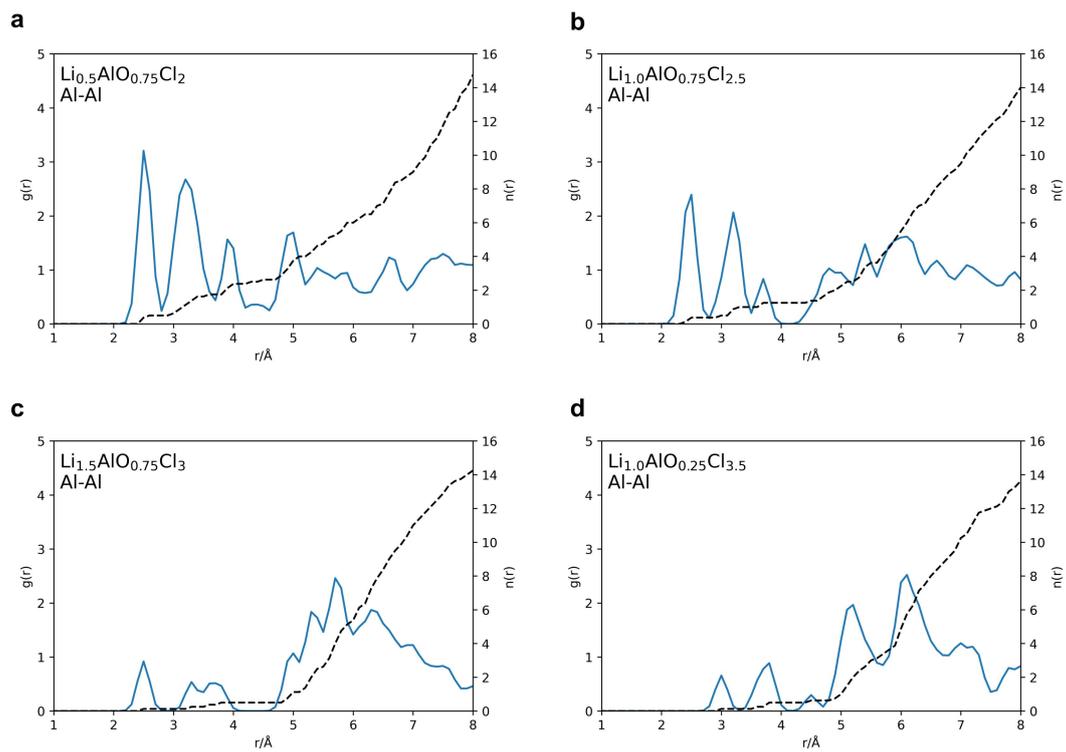

**Figure S4.** The RDF curves for Al-Al pairs of amorphous LAOCs at 300K.

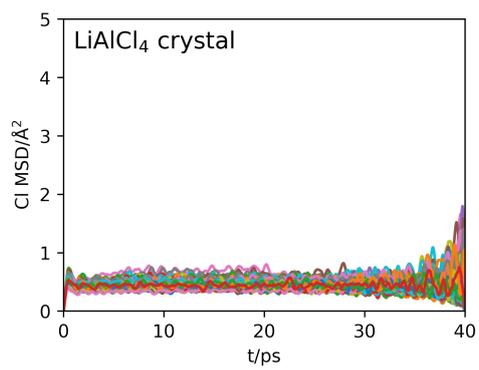

**Figure S5.** MSD of each Cl atom for crystal LiAlCl$_4$ calculated by AIMD at 300K.

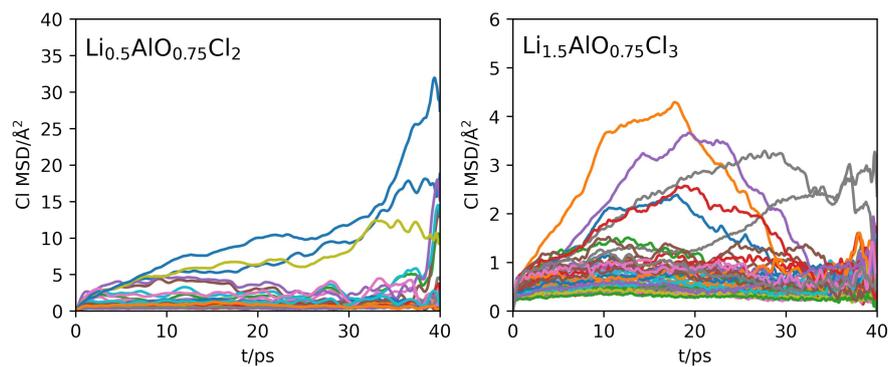

**Figure S6.** MSD of each Cl atom for $Li_{0.5}AlO_{0.75}Cl_2$ and $Li_{1.5}AlO_{0.75}Cl_3$ calculated by AIMD at 300K.

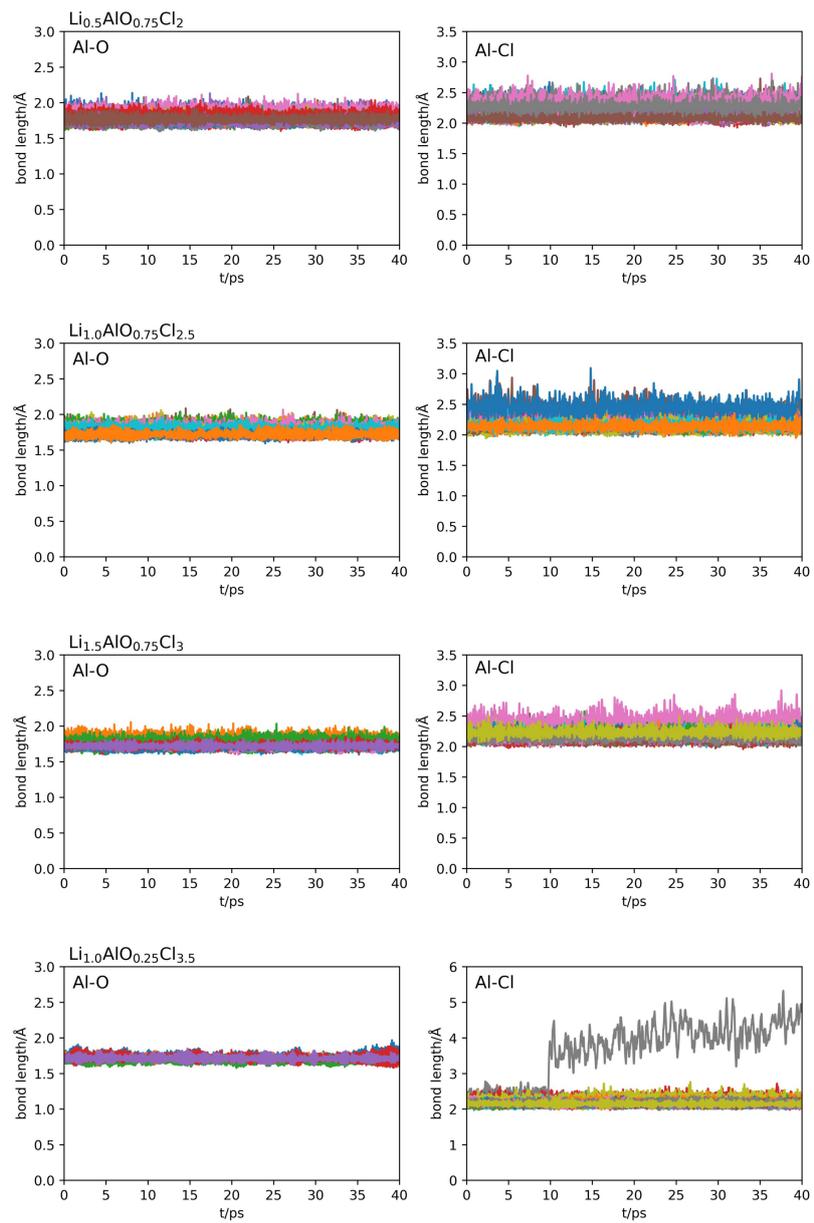

**Figure S7.** The Al-O and Al-Cl bond length over AIMD simulation at 300K of LAOCs.

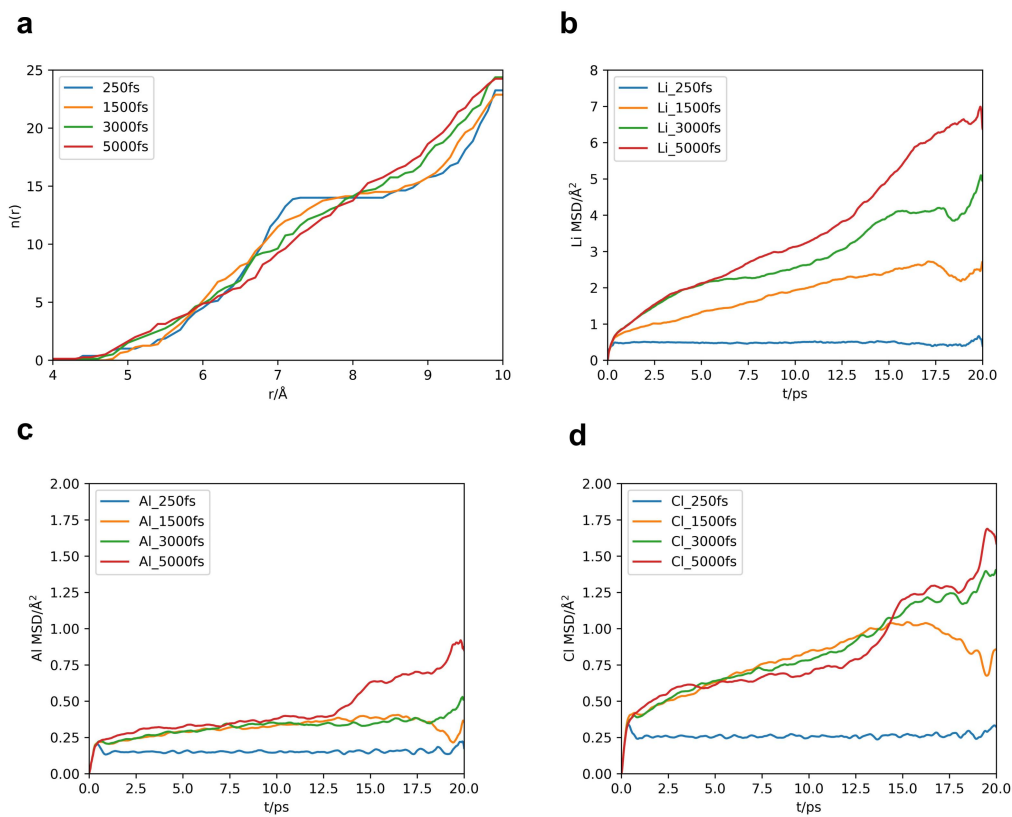

**Figure S8.** (a) The coordination numbers integrated from RDFs for Al-Al pairs of the LiAlCl$_4$ structures with different melting times. (b) The Li MSDs of the LiAlCl$_4$ structures with different melting times at 300K within the range of 11-50ps. (c). The Al MSDs of the LiAlCl$_4$ structures with different melting times at 300K within the range of 11-50ps. (d). The Cl MSDs of the LiAlCl$_4$ structures with different melting times at 300K within the range of 11-50ps.

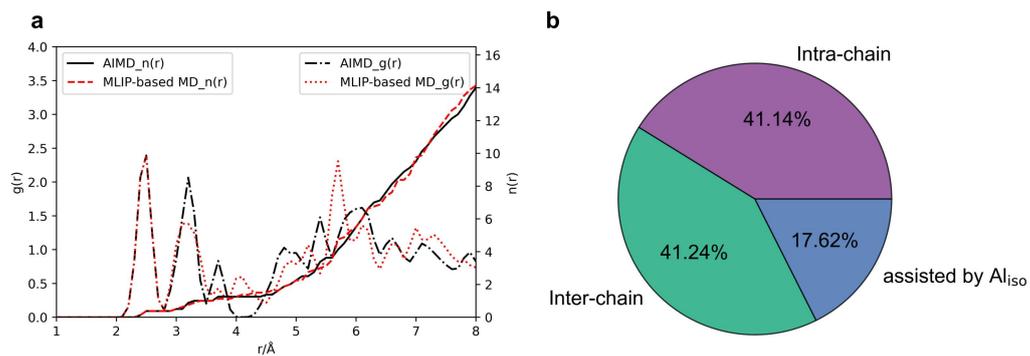

**Figure S9.** (a). The Al-Al RDF and coordination number curves of the $Li_{1.0}AlO_{0.75}Cl_{2.5}$ original cell by AIMD and MLIP-based MD. (b). The percentage of three different transport events in $Li_{1.0}AlO_{0.75}Cl_{2.5}$ with MLIP-based MD simulation at 300K within the range of 11-50ps.

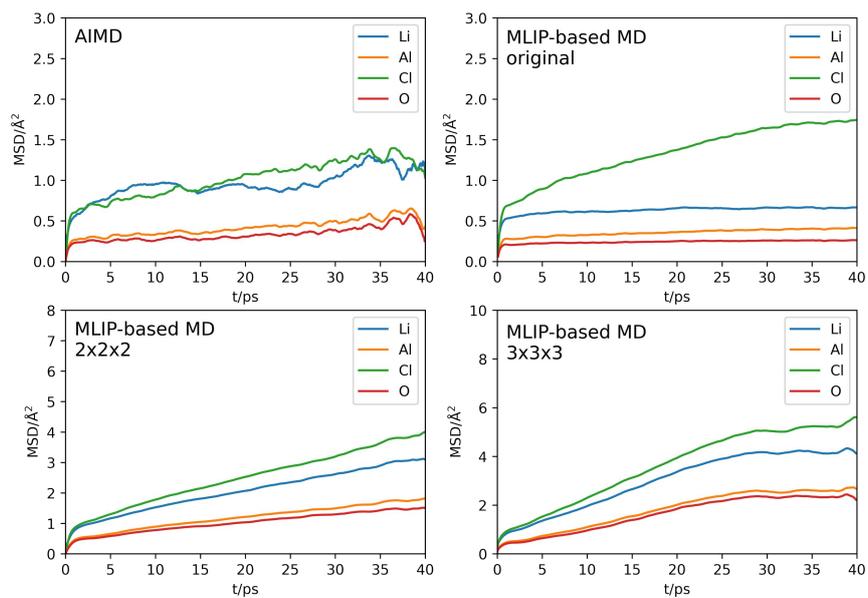

**Figure S10.** The MSDs at 300K of $Li_{1.0}AlO_{0.75}Cl_{2.5}$ by AIMD and different cell sizes by MLIP-based MD.

**Table S1.** Total number of atoms in cells for AIMD simulations.

| Structure | Total Number of Atoms in Cell |
|---|---|
| crystal LiAlCl$_4$ | 96 |
| Li$_{1.0}$AlCl$_4$ | 96 |
| Li$_{0.5}$AlO$_{0.75}$Cl$_2$ | 68 |
| Li$_{1.0}$AlO$_{0.75}$Cl$_{2.5}$ | 84 |
| Li$_{1.5}$AlO$_{0.75}$Cl$_3$ | 100 |
| Li$_{1.0}$AlO$_{0.25}$Cl$_{3.5}$ | 92 |

**Table S2.** Total number of atoms in cells of $Li_{1.0}AlO_{0.75}Cl_{2.5}$ for MLIP-based MD simulation.

| Structure | Total Number of Atoms in Cell |
|---|---|
| original cell | 84 |
| 2×2×2 cell | 672 |
| 3×3×3 cell | 2268 |

**Note S1. Verification of the accuracy of MLIP model and the study on the size effect.**

Firstly, the accuracy of this model was evaluated. AIMD simulations and MLIP-based MD simulations of the $Li_{1.0}AlO_{0.75}Cl_{2.5}$ original cell at 300K were conducted respectively for evaluating the rationality of the model. The Al-Al RDF of each structure is adopted to analyze the similarity among structures. As we can see in Figure S9a, the Al-Al RDF and coordination number curves of the two structures, especially the first nearest neighbor coordination, are very similar, so it can be considered that their structures are basically homologous. In addition, the accuracy of the model can be verified by comparing the statistical results of ion transport events between two different structures. Statistical results in Figure S9b is very similar to that of $Li_{1.0}AlO_{0.75}Cl_{2.5}$ in Figure 2b, indicating the two simulation processes reveal the same migration characteristics. Additionally, the MSDs in AIMD simulation and MLIP-based MD are similar in original cell, which further proves the reliability of the model. Therefore, the effectiveness of the MLIP model is demonstrated from three aspects: structure, migration events, and MSD. Secondly, the size effect of the MLIP model was explored by conducting MLIP-based MD simulation of the $Li_{1.0}AlO_{0.75}Cl_{2.5}$ original cell, the expanded 2×2×2 and 3×3×3 cell at 300K. By Figure S10, it can be deduced that the MSDs of 2×2×2 and 3×3×3 cell already formed a good convergence, so we chose 3×3×3 cell to conduct more MLIP-based MD simulations for the following analysis.

**Note S2. Simulation Process of Model 1 and Model 2.**

There are two different model building methods: Model 1 refers to MLIP-based MD calculations of a process of re-melting for 300ps at 600K, quenching for 3ps from 600K to 300K, and equilibrating for 500 ps at 300K; Model 2 refers to the direct conduction of MLIP-based MD simulations on the expanded structure for 500 ps at 300K.